# Could life have been transferred from Mars to Earth? Laboratory and computational simulations of Martian ejecta.

*Gregory M Davis[1,2*], Jonathan Horner[1], Bradley D Carter[1] and Stephen C Marsden[1]*

[1] *Centre for Astrophysics, University of Southern Queensland, Toowoomba, QLD 4350, Australia.*

[2] *Department of Health Sciences and Biostatistics, Swinburne University of Technology, Hawthorn, Australia.*

**Summary:** The study of the origin of life on Earth has been broadened due to panspermia models that suggest that early life may have been transferred between planets. Mars likely once had conditions that could support life, and it is interesting therefore to consider the question of early interplanetary transfer of life from Mars to the Earth. Endospore forming bacteria are ideal candidates for these studies as they can withstand harsh environmental conditions. For this reason, the idea that early life could have been delivered to Earth on Martian ejecta in the late Hadean period has gained considerable interest. To assess this, we have performed a series of both biological and astrophysical experiments. We exposed endospores shielded by a lysed colony of bacteria to extended UVC irradiation under a variety of rotation regimes, to simulate interplanetary exposure on ejecta with a variety of rotation periods. We also performed detailed n-body simulations of particles ejected from Mars at both perihelion and aphelion, finding that Martian ejecta can reach the Earth on timescales of just a few years - suggesting that, with ejection at the ideal time, transfer could occur within one year. Taken together, this study suggests this interplanetary transfer of biologically viable material from Mars to Earth is plausible under favourable conditions.

## Introduction

For generations, humanity has wondered where we came from. How did life begin on Earth and are we alone in the cosmos? In recent years, our knowledge of the earliest life on Earth has grown significantly, with ancient fossils revealing that life was already present on our planet during the late Hadean or early Archean periods, between 3.7 and 4.2 Ga [1, 2]. Indeed, it seems likely that life was present on Earth almost as soon as conditions on our planet allowed life to develop and thrive. Despite this, the true origin of life on Earth remains unknown. What we do know, however, is that the early Solar system was a remarkably violent place. Our planetary system formed approximately 4.6 Gyr ago. The growth of the terrestrial planets occurred over a period of tens of millions of years, with the latter stages of that formation marked by the occurrence of giant collisions between planet-sized bodies (see [3] and references therein for a broad overview of our knowledge of the Solar system's formation, and [4] for a review of our knowledge of the first five million years of planet formation in the Solar system). A key moment in this cleanup phase of terrestrial planet formation was the giant Moon-forming impact, where the Earth is thought to have collided with a Mars-sized object often referred to as Theia [5, 6]. Even if life were present on the Earth before that impact, that collision would have sterilised our planet, rendering it inhospitable for several million years. This giant impact suggests that the Earth was likely the last of the terrestrial planets to form [7]. A growing body of studies suggests that life could be transferred between planets through a process known as 'panspermia', where impact events on one planet eject rocky material that travels through interplanetary space before landing on another planet (see e.g. [43][44][45][46][47][48][24][25]). The theory of 'panspermia' is that life could have potentially been transported on such material between the planets.

Recent understanding of Mars has supported interest in panspermic theories that suggest that life may have arrived on Earth from Mars via interplanetary ejecta [8]. This stems from the idea that life may have evolved on Mars, then reached Earth on Martian ejecta in the late Hadean period. Although such panspermia theories are viewed with scepticism, they do hold merit when considering the early topography and atmosphere of Mars. It is widely accepted that Mars once had abundant surface water - with clear evidence of riverbeds, lakes, river deltas, and even suggestions that the low-lying northern hemisphere of Mars was once covered by a deep ocean [3, 9-11]. Current rover missions have unveiled not only the current chemical composition of several regions of Mars, but have also identified several signatures preserved in Martian regolith that suggest signs of early life [12-15]. The most promising of these being the precursors of fatty acids, which are required for cell membrane construction [16]. Although these can form in non-biological situations, this discovery is a significant leap in determining if Mars once harboured simple life forms and gives merit to panspermia theories that rely on Mars as the origin of life. Some bacteria can form endospores, which protect bacteria from damaging environmental conditions such as ultraviolet radiation (UV), toxins and chemicals that would otherwise disrupt survival [17]. An ideal candidate bacteria for observing this is *Bacillus subtilis*, which is one of the most well understood endospore forming bacteria and commonly used in astrobiology studies [18]. Examples of this include their use in the EXPOSE-E and EXPOSE-R missions [19, 20]. Additionally, these endospore formers have also been shown to be able to withstand the pressures and shocks that would be sustained in an ejecta event [21, 22], making them ideal targets for the study of whether life could be transferred from Mars to the Earth.

Hypothetical Martian ejecta would face considerable hurdles on a voyage from Mars to Earth. These include radiation in the form of UVC, solar wind and cosmic rays which would breakdown endospores exposed on the surface of the host meteoroid [23]. Ejecta moving through interplanetary space display a wide variety of rotation periods, from ultra-fast (rotation periods of seconds or minutes) to ultra-slow (periods of hundreds or thousands of hours), which would likely have an impact on the long-term viability of the bacteria that ejecta transports. We have previously shown that fast rotating simulated Martian regolith with endospores exposed to UVC in laboratory conditions withstand degradation for longer than slower rotators [24]. In addition to this, we have previously reported that lysed bacteria extend this protection from UVC if endospores were to form under a colony of exposed bacteria [25]. Slower rotating debris, with rotation periods of several hours or longer, would result in surface bacteria experiencing prolonged periodic exposure of UVC [26, 27]. This longer period of exposure is also coupled with longer periods of sheltering which, if partially protected, could allow time for bacterial DNA to initiate repair mechanisms.

In this study, we employ a biological and astrophysical approach to exploring long exposures of partially protected *B. subtilis* endospores. From a biological perspective, we exposed endospores to long periodic exposures of UVC which being partially protected by lysed bacteria and found a similar trend to our previous findings involving faster rotators. In addition to this, we performed astrophysical simulations of the orbital evolution of test particles ejected from Mars to examine the timescales upon which Martian debris can be delivered to the surface of the Earth. Our simulations revealed that debris ejected from Mars can be rapidly transferred to Earth, with some particles reaching our planet within just a few years of their ejection from Mars. Coupled together, our work suggests that life could theoretically survive the journey from Mars to Earth, if given the most favourable conditions.

# Methods

**Simulations of Martian ejecta**

To determine the range of timescales on which ejecta from Mars reaches the Earth, we performed four suites of detailed *n*-body simulations using the Hybrid integrator within the *n*-body dynamics package Mercury [28]. The version of the code used was modified to include the effects of the first order post-Newtonian corrections, following [29-31]. The orbits and masses of the Solar system planets in those simulations were obtained from the Horizons DE431 ephemeris. A small pre-simulation run was carried out to integrate the orbits from that ephemeris forward in time for one full orbit of Mars, with data output on an hourly basis.

Using the output from that pre-simulation run, the instantaneous cartesian orbital elements for all eight planets were obtained for the hour at which Mars was located at perihelion, and when it was located at aphelion. These two sets of orbital elements formed the initial conditions for our suites of *n*-body simulations to study the transfer of ejecta from Mars to Earth. Two different scenarios were considered for the location of Mars at the instant of a collision that ejected material from Mars – ejection at perihelion, and ejection at aphelion. For each of these scenarios, two different sub-scenarios were considered. In the first, the ejecta was launched forward of Mars - being ejected in the direction of motion of Mars at that instant. In the second, the eject was launched backward from Mars - launched along the velocity vector of Mars, but moving slower than Mars around the Sun - as though it had been ejected in the opposite direction to Mars' motion.

The Earth is an incredibly small target in space, and so to ensure that a reasonable number of impacts on our planet were recorded within our simulations, we followed the same methodology as laid out in [32-35]. Specifically, the radius of the Earth in our simulations was artificially inflated to be one million kilometres. This criterion was developed in [32], in order to obtain a statistically significant sample of impacts within an integration timescale that was humanly achievable, and significant amounts of benchmarking work was carried to confirm that the inflation of the Earth increased the impact rate in the expected manner. That benchmarking demonstrated that the number of impacts occurring on Earth scale directly with the cross-sectional area of our planet.

In each simulation, 100,000 massless test particles were placed at an initial location at a distance of one Martian Hill Radius ahead (or behind) Mars, with initial velocities in the same direction as Mars' motion, travelling at between 0.001 and 8.000 kilometres per second with respect to Mars (equivalent to ejection speeds at Mars' surface between 5.028 km/s and 13.027 km/s; noting that Mars' escape velocity is 5.027 km/s), following the ejecta velocities detailed in [36]. The orbital evolution of those 100,000 particles was followed for a period of 10 Myr under the gravitational influence of the eight planets and the Sun. Particles were removed from the system upon collision with one of the planets or the central body, or when they reached a heliocentric distance of 100 au. The time at which particles were removed from the system was recorded, giving a dataset that details the times at which particles collided with the inflated Earth, forming the basis for our dynamical results described below.

### Bacterial strains and cultivation of *B. subtilis* endospores

UVC irradiation assays utilised *B. subtilis* FUHAC10 and *S. aureus* ATCC25923 as a negative control (both strains were kindly provided by Andrew R. Greenhill, Federation University). These were grown overnight at 37 °C in nutrient broth (ThermoFisher), then plated and maintained on nutrient agar (NA) plates. Negative controls were applied as vegetative cells and *B. subtilis* endospores were generated using the methods as described in [37] and endospores were purified using the wash method as outlined in [25]. Endospores were assessed for viability on NA plates overnight at 37°C. Lysed bacteria which produced a shield for endospores exposed to UVC radiation were generated by centrifuging a slurry of pre-irradiated *B. subtilis* culture that was non-viable at 10,000g for 10 minutes at room temperature. Excess liquid was removed until an optical density reading of OD600 = 1.5 was achieved via serial dilutions using a Implen OD600® (Implen) photometer which represents 1.5 x $10^8$ cells/mL.

### Periodic UVC exposures of *B. subtilis* endospores

Irradiation of endospores was achieved by exposing *B. subtilis* endospores using a Gelaire BHEN biosafety cabinet (Gelaire) with an average wavelength of 245nm. For each timepoint, 50µL of purified endospores were placed on the surface of 20 empty 35mm agar plates, then covered with pre-lysed bacteria to create a covering of 10µm which was calculated at 7.07µL per dish. Exposures were timed and after removal, the wash plate method was used to transfer any viable endospores to NA plates and incubated overnight. Viability was determined based on successful growth from each plate rather than on a cellular basis. Each time point assay was repeated 3 times and each assay ran for 16 days.

## Results and Discussion

### Dynamical Simulation Results

Since the orbit of Mars is eccentric (e ~0.0934), the orbital velocity of the planet varies somewhat through the course of each orbit. As a result, the initial orbits of test particles upon escaping from Mars' Hill sphere will vary markedly depending on both the ejection location in Mars' orbit, and the ejection geometry. To address this, we considered four distinct scenarios - ejection at perihelion, in the forward direction; ejection at perihelion in the backward direction, ejection at aphelion in the forward direction, and ejection at aphelion in the backward direction. The instantaneous orbits attained by the ejecta would have greater semi-major axes than the orbit of Mars for ejection in the forward direction, meaning that no particles ejected in that manner would be emplaced on Earth-crossing orbits. By contrast, objects ejected in the backward direction would move on initial orbits with semi-major axes smaller than that of Mars, meaning that at least a subset of those ejecta would move on Earth-approaching or Earth-crossing orbits. By choosing the four scenarios we studied, we attempted to take the most extreme possible scenarios to give us a basis to calculate the typical spread of transfer times for Martian debris to the Earth.

As expected, the ejection direction of the test particles had a significant impact on the outcomes for the ejecta (Table 1). Ejection at perihelion in a backwards direction resulted in a greater number of impacts with Earth than perihelion backwards, with those ejected forwards having significant numbers of particles ejected from the system or colliding with the Sun and other planets. The simulations featuring ejection at Mars' aphelion lead to results that were very similar to those featuring ejection at perihelion. Across the two suites of simulations, it is clear that ejection in the backwards direction results in more impacts on Earth than ejection in the forward direction from Mars - exactly as one might expect from consideration of the initial orbits on which that ejecta will be placed.

*Table 1: The number of test particles that were ejected or that collided with one of the planets or the Sun through the course of our test simulations. In those simulations, the radii and masses of all the planets and the Sun were set to their canonical values, with the exception that the Earth was inflated to have a radius of one million kilometres, in order to ensure a statistically significant sample of impacts to determine the temporal spread of material transferred between Mars and the Earth. The four scenarios tested were: Ejection in a forward direction at Mars' perihelion; ejection in a backwards direction at Mars' perihelion; forwards ejection at Mars' aphelion; and backwards ejection at Mars' aphelion. The row labelled 'Equivalent Collisions at 1 $R_\oplus$' is an estimate of the number of collisions that would have been experienced by the Earth in these simulations if the radius of Earth in the simulations were the true value; these values were obtained by scaling the collisions counted in our simulations by a factor of 24,500.*

| **Outcome** | Perihelion Forwards | Perihelion Backwards | Aphelion Forwards | Aphelion Backwards |
|---|---|---|---|---|
| **Collision with the Sun** | 713 | 0 | 314 | 0 |
| **Collision with Mercury** | 0 | 0 | 0 | 0 |
| **Collision with Venus** | 2 | 4 | 0 | 3 |
| **Collision with the inflated Earth [Equivalent Collisions at 1 $R_\oplus$]** | 37299<br>1.52 | 86635<br>3.54 | 35203<br>1.44 | 88231<br>3.60 |
| **Collision with Mars** | 5268 | 5470 | 4379 | 3816 |
| **Collision with Jupiter** | 445 | 0 | 457 | 0 |
| **Collision with Saturn** | 33 | 0 | 34 | 0 |
| **Collision with Uranus** | 1 | 0 | 0 | 0 |
| **Collision with Neptune** | 0 | 0 | 1 | 0 |
| **Ejection from the system** | 34483 | 0 | 27089 | 0 |

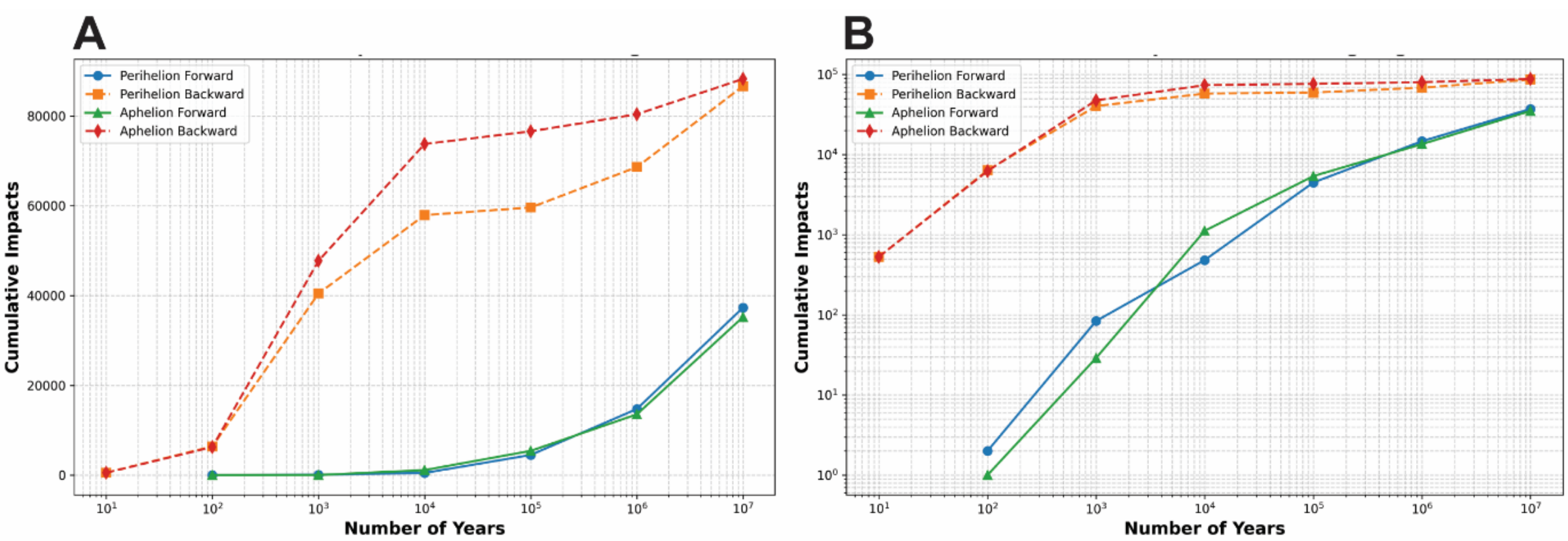

***Fig.1: Number of impacts upon the inflated Earth over time through our four simulated scenarios:*** Number of impacts on a log-linear scale (A), and the same data represented on a log-log scale (B). The impact of the ejection direction is immediately apparent, with particles

ejected behind Mars (initially travelling slower than Mars in its orbit) reaching Earth far more rapidly, and causing far more impacts, than ejecta flung ahead of Mars in its orbit.

From all four simulated scenarios there were test particles that collided with the Earth as a function of time (Fig.1). Ejections from Mars in the backward direction lead to a very rapid instigation of impacts on Earth - with the first collisions on our planet happening within the first decade after the particles are ejected from Mars. Impacts from ejecta flung in the forwards direction from Mars are somewhat delayed compared to those from backward-directed ejection – a direct result of the initial orbits on which the ejecta are emplaced. For ejection behind Mars in its orbit, the ejecta will be placed on orbits that have the potential to be immediately Earth-crossing, whilst those ejecta flung forward of Mars in its orbit must undergo significant dynamical evolution before being placed on Earth-crossing orbits. Taken together, our results reinforce the idea that material is readily transported between Mars and the Earth as a natural outcome of impacts by asteroids and comets on the Martian surface. They clearly demonstrate that, across the age of the Solar system, significant quantities of Martian material will have been transported to our planet whilst only spending a few months, or a few years, exposed to the hard vacuum and hard radiation environment if interplanetary space.

**Periodical exposure of UVC influences endospore viability**

Celestial bodies arising from hypothetical ejecta from Mars harbouring endospores would be exposed to UVC radiation, and the frequency of this dosage would be relative to their rotational frequency. We have previously established that endospores can survive longer UVC exposures if shielded by lysed bacteria which represent lysed members of their colony [25]. In addition to UVC absorbance, we have additionally established that *B. subtilis* endospores are more sensitive to faster pulsing of UVC which represents fast rotators [24]. Therefore, in order to address ejecta that rotated slower, but partially shielded from UVC radiation, *B. subtilis* endospores were subjected to longer exposures of UVC radiation. Slow rotators have been described as being a range of 1- to 20 hours, but can be as slow as ~200 hours [2, 3]. Due to this, for simplicity and resources, exposure times were set for 6, 12, 24 and 48 hour intervals and ran for a period of 16 days. Although 16 days is not near the time required for a slow rotating meteor to reach Earth from Mars, it does give an indication regarding feasibility of endospore survival when partially shielded from UVC.

Since we had already addressed the viability of endospore survival on different variants of Martian regolith [24], these exposures were carried out on empty, sterile laboratory grade petri dishes, then covered with 10µm of lysed bacteria. This lysed slurry consisted of lysed spores and vegetative cells and its optical density was determined prior to exposures to provide an adequate shield. Prior to irradiation, the effective dose needed to be determined per hour, which was calculated for each timepoint (Table 2). In this manner, all timepoints were exposed for their associated time, then left to rest for the same duration. Despite the lack of rotation, which did not allow for partial shading that would occur during mid-rotation, this was representative of ejecta with longer rotational times. Previous studies have indicated that shorter pulsations of UVC are more detrimental to *B. subtilis* endospores than longer exposures, even when receiving the same UVC dose [24, 38, 39]. Interestingly, our findings suggest a similar trend (Fig.2), where 6 hour time intervals showed a faster rate of endospore decay than 48 hour exposures at the same UVC dosage. This was consistent when observing the time intervals of 12 and 24 hours, where 12 hour intervals were more sensitive when compared to 24 and 48 hour intervals respectively. This suggests that endospores, even when partially shielded, have a propensity to break down when the exposure is more frequent.

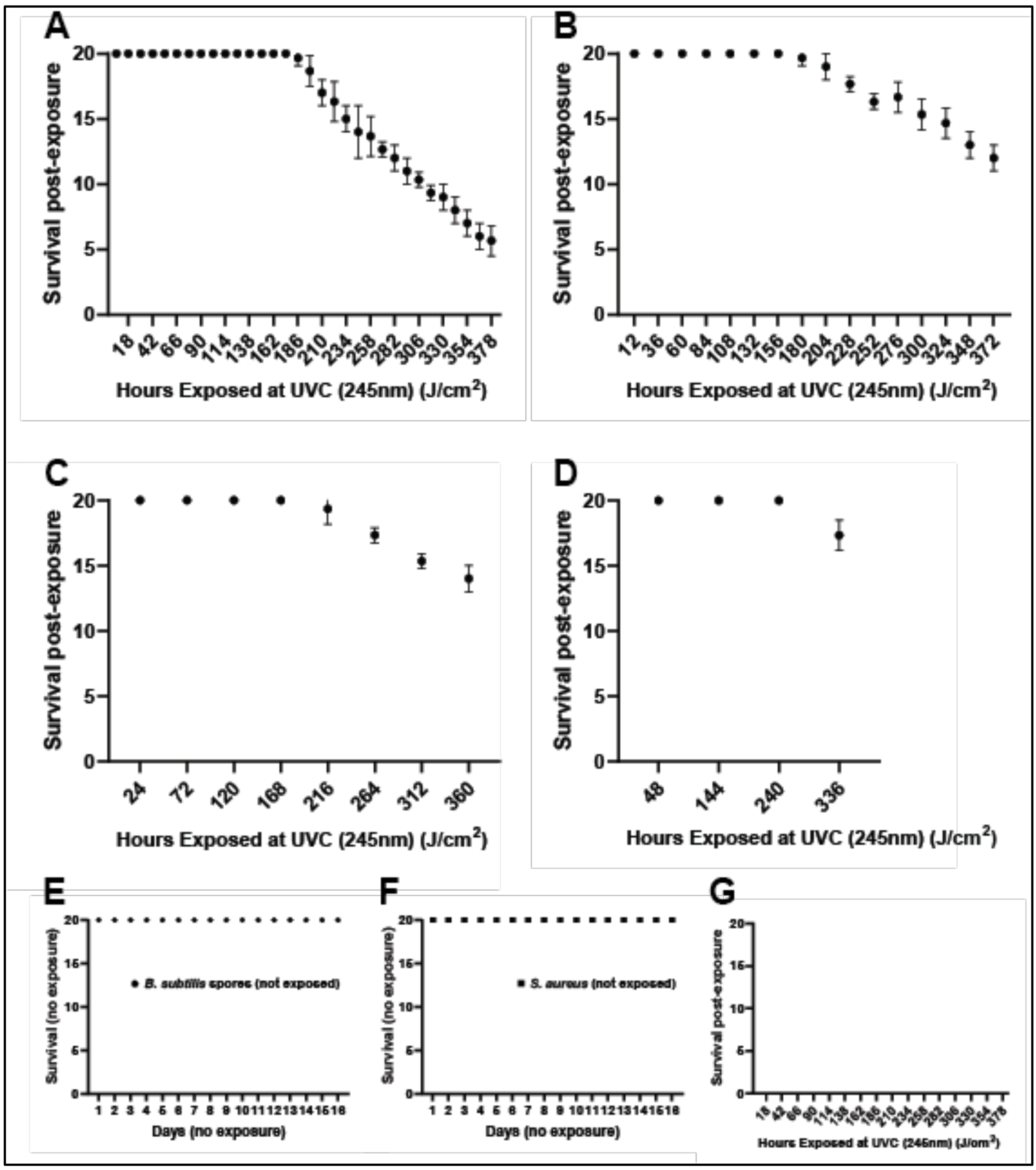

***Fig.2: Periodic exposures of B. subtilis endospores at different durations:*** *B. subtilis* endospores exposed to UVC at (A) 6 hour increments (B) 12 hour increments (C) 24 hour increments and (D) 48 hour increments. (E & F) Negative controls of unexposed *B. subtilis* endospores and *S. aureus* vegetative cells and (G) positive control of *B. subtilis* endospores and *S. aureus* vegetative cells exposed with each 6 hour timepoint showing no growth. Exposure relates to 20 samples removed each time point per exposure (Y-axis) vs Hours of exposure (X-axis). Error bars represent standard error of the mean. Exposures were performed in triplicates (N=60).

*Table 2: UVC dose (UVC J/cm2) per timepoint (hours).*

| 6 Hours | (UVC J/cm$^2$) | 12 hours | (UVC J/cm$^2$) | 24 hours | (UVC J/cm$^2$) | 48 hours | (UVC J/cm$^2$) |
|---|---|---|---|---|---|---|---|
| **6** | 24.42 | **12** | 48.84 | **24** | 97.68 | **48** | 195.36 |
| **18** | 48.84 | **36** | 97.68 | **72** | 195.36 | **144** | 390.72 |
| **30** | 73.26 | **60** | 146.52 | **120** | 293.04 | **240** | 586.08 |
| **42** | 97.68 | **84** | 195.36 | **168** | 390.72 | **336** | 781.44 |
| **54** | 122.1 | **108** | 244.2 | **216** | 488.4 | | |
| **66** | 146.52 | **132** | 293.04 | **264** | 586.08 | | |
| **78** | 170.94 | **156** | 341.88 | **312** | 683.76 | | |
| **90** | 195.36 | **180** | 390.72 | **360** | 781.44 | | |
| **102** | 219.78 | **204** | 439.56 | | | | |
| **114** | 244.2 | **228** | 488.4 | | | | |
| **126** | 268.62 | **252** | 537.24 | | | | |
| **138** | 293.04 | **276** | 586.08 | | | | |
| **150** | 317.46 | **300** | 634.92 | | | | |
| **162** | 341.88 | **324** | 683.76 | | | | |
| **174** | 366.3 | **348** | 732.6 | | | | |
| **186** | 390.72 | **372** | 781.44 | | | | |
| **198** | 415.14 | | | | | | |
| **210** | 439.56 | | | | | | |
| **222** | 463.98 | | | | | | |
| **234** | 488.4 | | | | | | |
| **246** | 512.82 | | | | | | |
| **258** | 537.24 | | | | | | |
| **270** | 561.66 | | | | | | |
| **282** | 586.08 | | | | | | |
| **294** | 610.5 | | | | | | |
| **306** | 634.92 | | | | | | |
| **318** | 659.34 | | | | | | |
| **330** | 683.76 | | | | | | |
| **342** | 708.18 | | | | | | |
| **354** | 732.6 | | | | | | |
| **366** | 757.02 | | | | | | |
| **378** | 781.44 | | | | | | |

**Biological and astrophysical constraints for successful transfer of life**

Our dynamical simulations reveal that material ejected from Mars can reach Earth on relatively short timescales. Indeed, significant amounts of Martian ejecta can reach Earth within just a few years or a few decades of their ejection. As a result, given the vast number of impacts on Mars since the formation of the Solar system, it is certain that significant amounts of Martian matter has transited to fall on our planet within a decade of its ejection from Mars. It is therefore important to consider the viability of bacteria that have spent only months or a few years in interplanetary space, rather than being concerned with residence times in orbit measured in thousands or millions of years. Simply put - if even a small number of bacteria can remain viable in vacuum, exposed to the interplanetary environment, for just a few years, then it is highly likely that any life from Mars would be able to inoculate the Earth.

The robust nature of *B. subtilis* endospores have shown them to remain viable in several harsh environments, including the vacuum of space [19, 20]. However, there are certain variables that would need to be factored in for a successful pollination of Earth from Martian ejecta. First of all, the UVC exposure in space is far greater than we can achieve in a laboratory simulation over the time period that our data is based on [40]. In addition to this, other forms of space weathering such as the solar wind and cosmic rays would need to be considered to represent the overall amount of radiation endospores would receive [23]. This further suggests that the partial protection that we show of endospores from lysed bacteria would be unlikely to provide adequate protection on the surface of Martian ejecta. Therefore, although it appears that bacterial survival on the surface of these ejecta would be problematic, survival of endospores embedded deeper within the surface of such ejecta would be feasible. This is based on the UVC penetrance of materials such as silicate or carbonaceous rock, which have greater protective penetrance depth against UVC radiation [41, 42]. Although organic material initially offers protection, our laboratory exposures suggest a breakdown of this protective organic layer over short time periods. Due to this, sub-surface endospores would receive greater protection and have a greater chance of being able to withstand these harsh conditions.

Despite the obstacles facing endospores in this type of environment, successful transfer of life between Mars and a primordial Earth is plausible. This does require factoring in the best chance for this planetary transfer to occur, such as the shortest potential transit time while enlarging the Earth to capture more potential impacts, and a large number of ejecta to overcome close encounters or collisions with other planets. Survival is enhanced if coupling this with the best conditions that would ensure the viability of endospores in space to survive otherwise harsh UVC exposure. Collectively, our n-body simulations, coupled with our biological data suggests that a planetary transfer of endospores from Mars to Earth, although challenging, is technically possible.

## Conclusion

The theory that life evolved on Mars and arrived on a primordial Earth, although heavily speculated, requires the successful transfer of simple life from Mars to Earth to gain plausibility. Our biological and astrophysical approach to this suggests that although challenging, the timespan and certain environmental constraints would allow the successful transfer of life from Mars to Earth.